\documentclass[aps,reprint,amsmath,amssymb,groupedaddress,article]{revtex4-1}
\usepackage{graphicx}
\usepackage{bbm}
\usepackage{verbatim}

\begin{document}

%Title of paper
\title{General quantum error-correcting code with entanglement based on codeword stabilized quantum code}

\author{Jeonghwan Shin}
\email{jhsh@korea.ac.kr}
\affiliation{School of Electrical Engineering, Korea University, Seoul, Korea}
\author{Jun Heo}
\email{junheo@korea.ac.kr}
\affiliation{School of Electrical Engineering, Korea University, Seoul, Korea}
\author{Todd A. Brun}
\email{tbrun@usc.edu}
\affiliation{Communication Sciences Institute, University of Southern California,
Los Angeles, CA 90089, USA}

\begin{abstract}
In this paper, we introduce a unified framework to construct entanglement-assisted quantum error-correcting codes, including additive and nonadditive codes, based on the codeword stabilized framework on subsystems.  The codeword stabilized (CWS) framework is a scheme to construct quantum error-correcting codes (QECCs) including both additive and nonadditive codes, and gives a method to construct a QECC from a classical error-correcting code in standard form.  Entangled pairs of qubits (ebits) can be used to improve capacity of quantum error correction.  In addition, it gives a method to overcome the dual-containing constraint.  Operator quantum error correction (OQEC) gives a general framework to construct quantum error-correcting codes.  We construct OQEC codes with ebits based on the CWS framework.  This new scheme, entanglement-assisted operator codeword stabilized (EAOCWS) quantum codes, is the most general framework we know of to construct both additive and nonadditive codes from classical error-correcting codes.  We describe the formalism of our scheme, demonstrate the construction with examples, and give several EAOCWS codes.
\end{abstract}



\maketitle

%%%%%%%%%%%%%%%%%%%%%%%%%%%%%%%%%%%%%%%%%
\section{Introduction}
\label{Sec:introduction}

Quantum error correction (QEC) plays an important role in quantum information processing and communication.  Without QEC it is impossible to maintain a quantum state against the corrupting effects of decoherence for long enough to carry out nontrivial quantum computations or communication protocols.  Since Shor introduced a method to encode  information qubits into a highly entangled state \cite{PhysRevA.52.R2493}, the field of QEC has developed rapidly into a large and diverse field of study.

In Refs.~\cite{PhysRevA.54.1098, PhysRevLett.77.793} it was shown that quantum error-correcting codes (QECCs) can be constructed from classical binary linear codes that satisfy a dual-containing constraint.  Stabilizer codes \cite{Gottesman:1996ub} are a general framework to construct quantum codes analogous to classical additive codes.

%=====================================================
% CWS code 
Recently, a more general method to construct QECC was introduced.  It was shown that both additive and nonadditive quantum codes can be constructed from classical codes using the codeword stabilized (CWS) framework \cite{Cross:2009jo}.  CWS codes consist of a unique base state, specified by a word stabilizer group, and a set of word operators that produce the other basis states of the code from the base state.  For a CWS code in standard form, the base state a graph state \cite{PhysRevA.69.022316}, with stabilizer generators corresponding to each vertex of the graph.  Using these  stabilizer generators, all single-qubit Pauli errors acting on a codeword state can be transformed into errors consisting only of $Z$ and identity operators.  CWS codes therefore correspond to classical codes that can correct a particular set of binary errors induced by the word stabilizer.

The word operators produce a set of basis states that spans the code space of the CWS code.  From the associated classical code, word operators of a CWS code that can correct the given set of Pauli errors can be identified.  The same procedure can be done with a set of multi-qubit Pauli errors to construct codes with higher distances.  Therefore, a CWS code in standard form is specified by a graph, whose vertices correspond to the qubits of the codeword, and a classical binary code \cite{Cross:2009jo}.

%=====================================================
% EA QEC code 
In QEC, maximally entangled pairs of qubits (ebits) shared between the sender (Alice) and receiver (Bob) can be used to improve the parameters of quantum codes, such as the minimum distance and/or code rate \cite{Lai:2010ve}.  The use of ebits also allows stabilizer codes to be constructed from classical codes without the dual-containing restriction \cite{Brun20102006}.   In addition, Ref.~\cite{PhysRevA.84.062321} showed that QECCs based on the CWS framework, and having minimum distances greater than 3, can be constructed by using ebits.  Without shared entanglement, the highest value of minimum distance for a CWS code constructed so far, based on the ring topology, is less than 4.

%=====================================================
% Operator QEC
Operator quantum error correction (OQEC) \cite{PhysRevLett.94.180501, PhysRevLett.95.230504} is a more general scheme to construct QECCS.  In this scheme, quantum information can be encoded into either a subspace (as in a standard QECC) or into a subsystem.  OQEC unifies both passive error-avoiding schemes and active error correction.  CWS codes, including the stabilizer codes and entanglement-assisted QECCs (EAQECCs) described above, encode quantum information into a subspace.  Recently, however, there has been work on constructing EAQECCs and CWS codes on subsystems.  A theory of entanglement-assisted operator QECCs was developed in \cite{PhysRevA.76.062313} in the stabilizer formalism, and it was shown how to construct CWS codes on subsystems in \cite{PhysRevA.86.042318}.

%=====================================================
% Contributions of this paper
In this paper, we provide a new construction method for QECCs on subsystems based on the CWS framework, using shared ebits.  This formalism of entanglement-assisted operator CWS (EAOCWS) codes gives a unified scheme for QECCs, including both additive and nonadditive codes.  Since EAOCWS codes are based on the CWS framework, these codes are also specified by a graph state and a classical binary error-correcting code, and can correct a set of errors induced by the word gauge group of the code.  The word gauge group includes both stabilizer operators (that leave the base state unchanged) and gauge operators that act only on the noisy subsystem.  All Pauli errors can be transformed by word gauge operators into effective errors consisting only of $Z$ and $I$ operators.  In standard form, the word gauge group of a CWS code is generated by the stabilizer generators of the base state plus some additional gauge operators consisting only of $Z$ and $I$ operators.  By applying these additional gauge operators to the effective errors, the $Z$ operators located on certain qubits can be removed \cite{PhysRevA.86.042318}.

It is assumed that Bob's halves of the shared ebits do not suffer from errors, because these qubits do not pass through the channel.  However, the word operators have non-trivial operators acting on Bob's qubits, because the effective errors induced by the word gauge operators can have $Z$ operators on Bob's qubits.  We will show that these word operators are equivalent to operators that only act nontrivially on Alice's side, even though the effective errors can act on Bob's qubits as well \cite{PhysRevA.84.062321}.  Therefore, encoding can be done by acting only Alice's qubits---a necessary requirement for a useful code.  However, we will see that in some cases this will restrict us to using only a subset of the word operators, and therefore only a subcode of the classical binary code.

%=====================================================
% Organization
This paper is organized as follows.  In Sec.~\ref{Sec:Notation_background}, we give an overview of notation and background with respect to general error correcting codes: codeword stabilized quantum codes, entanglement-assisted quantum error correcting codes, and operator quantum error correction.  In Sec.~\ref{Sec:Entanglement-assisted operator CWS codes}, we describe the unified framework to construct entanglement-assisted quantum error correcting codes on subsystems within the CWS framework, and give some examples of EAOCWS codes.  Finally, in Sec.~\ref{Sec:Conclusions}, we conclude.

%% EAOCWS ==================================

%%%%%%%%%%%%%%%%%%%%%%%%%%%%%%%%%%%%%%%%%
\section{Notation and background}
\label{Sec:Notation_background}

%======================================================
\subsection{Codeword stabilized quantum codes}

Codeword stabilized (CWS) codes \cite{Cross:2009jo} are a broad class of quantum error-correcting codes that include both additive and nonadditive quantum codes.  Stabilizer codes can be considered a subset of CWS codes (though generally not in standard form).

Nonadditive codes and additive codes have a difference in the dimension of the code space.  An additive (stabilizer) code encodes a definite number $k$ of logical qubits into a codeword of $n$ physical qubits.  Such a code with minimum distance $d$ is denoted an $[[n,k,d]]$ code.  The dimension of the codespace is $K=2^k$.  For a nonadditive code, the dimension $K$ of the code space need not be a power of 2.  Thus we introduce a different notation for nonadditive codes; we denote a nonadditive quantum code that encodes a $K$-dimensional code space into $n$ physical qubits with minimum distance $d$ as an $((n,K,d))$ code.

Theorem 1 in \cite{Cross:2009jo} showed that a CWS code is locally Clifford-equivalent to a form specified by a graph $G$ and a classical binary code.  This is called standard form.  In standard form, the graph $G$ and its adjacency matrix $A$ determines the word stabilizer of the CWS code.  The graph $G$ has $n$ vertices, one for each qubit of the codeword.  The word stabilizer $\mathcal{S}$ is a maximal Abelian subgroup of the Pauli group $\mathcal{P}_n$, and has a set of generators corresponding to the vertices of the graph $G$.  For a CWS code in standard form, the codeword stabilizer generators $\{S_i\}$ have the following structure:
\begin{equation}
S_i=X_iZ^{\mathbf{r}_i} ,
\end{equation}
where $\mathbf{r}_i$ is the $i$th row vector of the adjacency matrix $A$.  That is, each generator $S_i$ has a Pauli $X$ operator on the qubit corresponding to vertex $i$ of the graph, Pauli $Z$ operators on the qubits corresponding to each of the neighbors of $i$, and identity operators $I$ on all the other qubits.  The word stabilizer $\mathcal{S}$ is generated by the set $\{S_i\}$.

A unique base state $|S\rangle$  is the common $+1$ eigenstate of the word stabilizer $\mathcal{S}$ specified by the graph $G$.  This state is fixed by any element $S \in \mathcal{S}$ of the word stabilizer:
\[
|S\rangle=S|S\rangle .
\]

The word operators $\{w_l\}$ are also elements of $\mathcal{P}_n$.  The code space is spanned by basis states obtained by applying word operators to the base state $|S\rangle$, and each basis state is of the form
\begin{equation}
|w_l\rangle=w_l|S\rangle.
\end{equation}
Therefore, the number of the word operators determines the dimension of the code space, and the word operators map the base state onto an orthogonal state.

Lemma 2 in \cite{Cross:2009jo} specified that any error in a correctible error set $\mathcal{E}$, acting on codewords of a CWS code in standard form, can be represented by another form consisting only of $I$ and $Z$ operators.  This equivalent error is called the {\it induced error}.  By multiplying the error by those word stabilizers that have $X$ operators at the same locations as the original error, all factors of $X$ in the correctible errors can be eliminated, leaving only $Z$ operators.  With the induced errors, we can map between the error set $\mathcal{E}$ and a set of classical binary errors.

The mapping $Cl_{G}(E_a)$ between the quantum error $E_a =Z^{\mathbf{v}}X^{\mathbf{u}}$ and a classical binary error is defined by
\begin{eqnarray}
\label{Eq:CWS_map} Cl_{G}(E_a = Z^{\mathbf{v}}X^{\mathbf{u}})=
\mathbf{v}\oplus\bigoplus_{l=1}^nu_l\mathbf{r}_l,
\end{eqnarray}
where $\mathbf{v}$ and ${\mathbf{u}}$ are binary vectors, $\mathbf{r}_l$ is the $l$th row vector of the adjacency matrix $A$ for ${G}$, and $u_l$ is the $l$th bit of $\mathbf{u}$.  Theorem 3 of \cite{Cross:2009jo} demonstrates the equivalence between the error correction of a CWS code in standard form and a classical code with this definition.  A CWS code in standard form, defined by a graph $G$ and a classical binary code $C_b$, detects errors from the set $\mathcal{E}$ if and only if $C_b$ detects errors from the set $Cl_{G}(\mathcal{E})$, and for each $E_a\in\mathcal{E}$,
\begin{eqnarray*}
\textrm{either}\phantom{1}Cl_{G}(E_a) &\neq& 0 , \\
\textrm{or, for each}\phantom{1}l,\phantom{1}Z^{\mathbf{c}_l}
E_a&=&E_a Z^{\mathbf{c}_l} ,
\end{eqnarray*}
where the $\mathbf{c}_l$ are the binary codewords from $C_b$.  So we can see that the word operators $w_l$ of the CWS code in standard form are derived from the codewords $\mathbf{c}_l$ of the binary code $C_b$ by
\begin{equation}
\mathcal{W} = \{ w_l \} =
\{Z^{\mathbf{c}_l}\}_{\mathbf{c}_l\in\mathcal{C}_b} .
\end{equation}

%======================================================
\subsection{Entanglement-assisted quantum error-correcting codes}

The rate of a quantum error-correcting code can be improved by using pairs of maximally entangled qubits (ebits) \cite{PhysRevA.66.052313}.  Entanglement also allows us to overcome the dual-containing constraint \cite{Brun20102006}.

It is convenient to explain the properties of a code over its initial code space (before the encoding operation), because initial code space is unitarily equivalent to the code space (after the encoding operation) by the unitary encoding operation $U$.

An $[[n,k,d;c]]$ EAQECC encodes $k$ logical qubits into $n+c$ physical qubits (including $c$ entangled pairs shared between Alice and Bob).  The initial state $|\psi'\rangle$ of the $[[n,k,d;c]]$ EA-QECC consists of $k$ information qubits $|\phi\rangle$, $m=n-c$ qubits in the state $|0\rangle$ and $c$ entangled states:
\begin{equation}
\label{eq:EA_initial} 
|\psi'\rangle_{EA}=|\Phi_+\rangle^{\otimes c}|0\rangle^{\otimes (n-k-c)}|\phi\rangle
\end{equation}
where $|\Phi_+\rangle$ is the maximally entangled state $\frac{1}{\sqrt{2}}(|00\rangle+|11\rangle)$ and it is shared by Alice and Bob.  The other qubits are all initially on Alice's side.

For the initial state $|\psi'\rangle_{EA}$, the stabilizer group is generated by stabilizer generators as follows:
\begin{equation}
\label{eq:EA_stabilizer} 
\begin{array}{cl}
Z_i |  Z_i, & \textrm{for}\phantom{1}i=1,\cdots,c, \\
X_j | X_j, & \textrm{for}\phantom{1}j=1,\cdots,c, \\
Z_i | I ,    & \textrm{for}\phantom{1} i=c+1,\cdots,n-k, \\
\end{array}
\end{equation}
where the operators on the left and right of the `$|$' act on the qubits on Alice's and Bob's sides, respectively.

The logical operators of the initial state $|\psi'\rangle_{EA}$ are
\begin{eqnarray*}
&Z_{n-k+1}| I&,\cdots,Z_n| I , \\
&X_{n-k+1}| I&,\cdots,X_n| I ,
\end{eqnarray*}
where all of the operators act on Alice's part. 

After an encoding operation $U=U_E \otimes I$, the encoded state is
\[
|\psi\rangle_{EA}=U(|\Phi_+\rangle^{\otimes c}|0\rangle^{\otimes (n-k-c)}|\phi\rangle\nonumber).
\]
The stabilizer generators of the encoded state $|\psi\rangle_{EA}$ are 
\begin{eqnarray*}
g_i=U_E Z_i^A(U_E)^\dagger| I ,\phantom{1} && \textrm{for}\phantom{1}i=c+1,\cdots,n-k, \\
g_j=U_E Z_j^A(U_E)^\dagger| Z_j ,\phantom{1}&& \textrm{for}\phantom{1}j=1,\cdots,c, \\
h_j=U_E X_j^A(U_E)^\dagger| X_j ,\phantom{1}&&\textrm{for}\phantom{1}j=1,\cdots,c,
\end{eqnarray*}
and the logical operators on $|\psi\rangle_{EA}$ are represented as follows: 
\begin{eqnarray*}
U_E Z_{n-k+i}U_E^\dagger| I , & \textrm{for}\phantom{1}i=1,\cdots,k , \\
U_E X_{n-k+i}U_E^\dagger| I , & \textrm{for}\phantom{1}i=1,\cdots,k .
\end{eqnarray*}
The stabilizer group $\mathcal{S}$ is generated by two subgroups: the isotropic subgroup $\mathcal{S}_I$ and the symplectic subgroup $\mathcal{S}_S$:
\begin{equation*}
\mathcal{S} = \langle\mathcal{S}_I,\mathcal{S}_S\rangle.
\end{equation*}
The symplectic subgroup is generated by $\mathcal{S}_S = \langle \{ g_1,...,g_c,h_1,...,h_c\} \rangle$, and the isotropic subgroup is generated by $\mathcal{S}_I = \langle\{g_{c+1},...,g_{n-k}\}\rangle$.
The minimum distance $d$ is equal to the minimum weight of the operators in $N(\mathcal{S}) - \mathcal{S}_I$.

%======================================================
\subsection{Operator code}

Operator quantum error correction (OQEC) \cite{PhysRevLett.94.180501, PhysRevLett.95.230504, PhysRevA.73.012340} generalizes the theory of quantum error correction (QEC) and gives a unified framework to construct both active error correction and passive error avoiding schemes such as decoherence-free subspaces and noiseless subsystems.   In OQEC, quantum information is encoded into a subsystem rather than a subspace.  Consider a fixed partition of a system's Hilbert space:
\begin{equation*}
\mathcal{H}=(\mathcal{A} \otimes \mathcal{B}) \oplus \mathcal{K} .
\end{equation*}
Here, the Hilbert space is partitioned into two subspaces, $\mathcal{K}$ and $\mathcal{A} \otimes \mathcal{B}$.  The subspace $\mathcal{A} \otimes \mathcal{B}$ is orthogonal to $\mathcal{K}$, and factors into two subsystems by the tensor product.

Quantum information can be encoded into subsystem $\mathcal{A}$ by preparing the information state $\rho^A$ in subsystem $\mathcal{A}$:
\begin{equation*}
\rho = \rho^A \otimes \rho^B \oplus 0^K ,
\end{equation*}
where $\rho^B$ is an any arbitrary state on the subsystem $\mathcal{B}$.  This subsystem is called the {\it noisy} or {\it gauge} subsystem; operations that affect only the gauge subsystem leave the encoded information unchanged.

%% Stabilizer formalism
It is possible to extend the stabilizer formalism to include OQEC codes \cite{PhysRevLett.95.230504}. In this case, we encode a state of $k$ logical qubits into $n$ physical qubits.  Let
$\mathcal{P}_n$ be the $n$-fold Pauli group.  The initial state before encoding can be represented by
\begin{equation}
|C\rangle=|0\rangle^{\otimes s}|\psi\rangle |\phi\rangle ,
\label{OQEC_canonical}
\end{equation}
where $|\phi\rangle$ is the $k$-qubit state we wish to encode into a subsystem, $|\psi\rangle$ is an arbitrary $r$-qubit state (which will correspond to the gauge subsystem), and the remaining $s=n-k-r$ qubits are ancillas in the state $|0\rangle$.  Even if $|C\rangle$ and $|C'\rangle=|0\rangle^{\otimes s}|\psi'\rangle|\phi\rangle$ are different (because $|\psi\rangle\not=|\psi'\rangle$), both states are considered to encode the same information. Therefore, $|C\rangle$ and $|C'\rangle$ are equivalent by a {\it gauge transformation}:
\begin{equation*}
|C\rangle=g|C'\rangle
\end{equation*}
where $g$ is an operator in the algebra generated by the {\it gauge group} $\mathcal{G}$.

The gauge group $\mathcal{G}$ of this OQEC code is a nonabelian subgroup of $\mathcal{P}_n$ generated by
\begin{equation*}
Z_{1},\dots,Z_{s+r},X_{s+1},\dots,X_{s+r}.
\end{equation*}
Defined in this way, the gauge group includes the stabilizer group of this code, $\mathcal{S}$, that is generated by $Z_1,\dots,Z_s$.

The algebraic structure of this trivial code carries over to the OQEC after encoding.  The initial state is encoded by a unitary operator $U$ in the Clifford group.  After encoding, the generators of the gauge group are $\{S_1,\dots,S_{s+r},g_{s+1},\dots,g_{s+r} \}$, where $S_i$ and $g_j$ are isomorphic to $Z_i$ and $X_j$ on the unencoded state:
\begin{equation}
S_i = U Z_i U^\dagger ,\ \ \ \
g_j = U X_j U^\dagger .
\end{equation}
With this definition of the gauge group, the error set $\mathcal{E}$ is correctable if and only if
\begin{equation}
E_a E_b \notin N(\mathcal{S})-\mathcal{G}
\end{equation}
for all $E_a, E_b\in\mathcal{E}$ \cite{PhysRevLett.95.230504}.  We characterize an operator code by the parameters $n$, $k$, and $d$ (just as for a standard stabilizer code), but also the number of
gauge qubits $r$; we write this as $[[n,k,r,d]]$.

%%%%%%%%%%%%%%%%%%%%%%%%%%%%%%%%%%%%%%%%%
\section{Entanglement-assisted operator CWS codes}
\label{Sec:Entanglement-assisted operator CWS codes}

In this section, we introduce a framework for entanglement-assisted CWS codes that encode quantum information into a subsystem. We call these entanglement-assisted operator codeword stabilized (EAOCWS) codes.  In an EAOCWS code, it is supposed that Alice and Bob share $c$ pairs of maximally entangled states, typically the Bell state $|\Phi_+\rangle$.  Furthermore, it is assumed that the halves of ebits held by Bob do not suffer from errors, since they do not pass through the noisy channel.  An EAOCWS code is defined  by a word gauge group $\mathcal{G}$ (including the word stabilizer $\mathcal{S}$), a base state $|S\rangle$, and word operators $\mathcal{W}$.

% 일반적인 cWS는 word stabilizer word operator unique state로 구성되어 있다. 이장에서 우리는 ebit을 이요한 cws가 subsystem에서 구성되기 위해서 각 요소 들이 어떻게 구성되는지 설명하고 실제 예를 통해서 EAOCWS를 보이겠다.

%% The initial unique state of EAOCWS code ==========================
%======================================================
\subsection{Initial base state of EAOCWS codes}

In a CWS code, the unique base state is defined by the word stabilizer.  Similarly, the base state of an EAOCWS code is specified by the word gauge group.  For convenience, we first consider the initial base state before applying the encoding unitary.  The initial base state $|S_b\rangle$ of the $((n,K,r,d;c))$ EAOCWS code consists of $s=n-r-c$ qubits in the state $|0\rangle$, $c$ ebits and $r$ gauge qubits in an arbitrary state:
\begin{equation}
\label{eq:initial_base_state}
|S_b\rangle=|0\rangle^{\otimes s}|\Phi_+\rangle^{\otimes c}|\psi\rangle
\end{equation}
where $|\psi\rangle$ is an arbitrary $r$-qubit state.  Each maximally entangled pair $|\Phi_+\rangle$ is shared by Alice and Bob, and it is assumed that the halves of ebits on Bob's side do not suffer from errors.  All other qubits are on Alice's side.

Because we are defining a code on a subsystem, the ``base state'' is not really a unique state; it is only defined up to the arbitrary state of the gauge subsystem.  We therefore identify an equivalence class of base states that can be turned into each other by gauge transformations $\mathcal{G}$:
\begin{equation}
|S_b\rangle \sim  |S_b'\rangle \phantom{1} \Leftrightarrow \phantom{1} |S_b\rangle = G|S_b'\rangle,
\end{equation}
where $G \in \mathcal{G}$.
This equivalence class of initial base states is stabilized by a stabilizer group $\mathcal{S}$, i.e.,
\begin{equation}
|S_b\rangle=S|S_b\rangle ,
\end{equation}
where $S \in \mathcal{S}$.

The minimal set of operators that can generate the gauge group of an EAOCWS code comprises three types of operators.  First, the word stabilizer $\mathcal{S}_b^s$ of the initial base state corresponds to a fixed $s$-qubit state in subsystem $\mathcal{A}$, and acts trivially on subsystem $\mathcal{B}$.  For the initial base state Eq.~(\ref{eq:initial_base_state}), the fixed state is $|0\rangle^{\otimes s}$, and the word stabilizer $\mathcal{S}_b^s$ is generated by the operators
\begin{eqnarray}
\label{Eq:word_stabilizer_initial}
\begin{array}{c}
Z_1II\cdots I|I^{\otimes c} \\
\vdots \\
I\cdots IZ_m I\cdots I |I^{\otimes c}\end{array} .
\end{eqnarray}
The operators on the left and right of the `$|$' act on the qubits on Alice's and Bob's sides, respectively.

The initial base state includes $c$ maximally entangled pairs of qubits between subsystem $\mathcal{A}$ and $c$ qubits on Bob's side.  The Bell state $|\Phi_+\rangle$ is stabilized by two operators $XX$ and $ZZ$.  Therefore, the word stabilizer $S_b^c$ acting on the $c$ ebits is generated by
\begin{eqnarray}
\label{Eq:word_stabilizer_EA_initial}
\begin{array}{c}
I \cdots IZ_{s+1}I\cdots I|Z_1 I \cdots I , \\
I \cdots IX_{s+1}I\cdots I|X_1 I \cdots I , \\
\vdots \\
I\cdots IZ_{s+c}I\cdots I| I \cdots I Z_c , \\
I\cdots IX_{s+c}I\cdots I| I \cdots I X_c .
\end{array}
\end{eqnarray}
The groups $S_b^s$ and $S_b^c$ stabilize the fixed state $|0\rangle^{\otimes s}|\Phi_+\rangle^{\otimes c}$ on subsystem $\mathcal{A}$.

Since the state $|\psi\rangle$ in the initial base state $|S\rangle_b$, Eq.~(\ref{eq:initial_base_state}), is an arbitrary state on subsystem $\mathcal{B}$, any operators acting only on subsystem $\mathcal{B}$ are elements of the gauge subgroup $S_b^g$ acting on $|\psi\rangle$.  Therefore, the gauge subgroup $S_b^g$ is generated by
\begin{eqnarray}
\label{Eq:word_stabilizer_gauge_initial}
\begin{array}{c}
I \cdots IZ_{n-r+1}I\cdots I|I^{\otimes c}, \\
I \cdots IX_{n-r+1}I\cdots I|I^{\otimes c}, \\
\vdots \\
\phantom{I\cdots I} I\cdots IZ_{n}|I^{\otimes c}, \\
\phantom{I\cdots I} I\cdots IX_{n}|I^{\otimes c}.
\end{array}
\end{eqnarray}

From the above three subgroups of the word gauge group, we generate the word gauge group $\mathcal{G}_b$ of the initial base state $|S_b\rangle$ for an EAOCWS code:
\begin{equation}
\mathcal{G}_b=\langle \{\mathcal{S}_b^s,\mathcal{S}_c^c,\mathcal{S}_b^g\}
\rangle
\end{equation}
The word gauge operators act trivially on subsystem $\mathcal{A}$ and leave the equivalence class of initial base states invariant.

%% The initial word operator of EAOCWS code
The code space is spanned by basis states given by applying the set of word operators $\mathcal{W}$ to the base state.  The encoding operation must be performed only  on Alice's side.  Therefore, the word operators act only on Alice's qubits as well.  For the initial base state $|S_b\rangle$, the initial word operators $\mathcal{W}_b=\{w'_l\}$ of an EAOCWS code can be represented as
\begin{equation}
\label{eq:base_word_operator}
w'_l=X^\mathbf{x} \otimes Z^\mathbf{u}X^\mathbf{w}\otimes
I^{\otimes r}|I^{\otimes c} ,
\end{equation}
where $\mathbf{x}$, $\mathbf{u}$ and $\mathbf{w}$ are binary vectors of length $m$, $c$ and $c$, respectively.  We could similarly define $w''_l$ having a non-trivial operators on subsystem $\mathcal{B}$ as a word operator for the initial base state.  The operator $w''_l$ is equivalent to $w'_l$ by a gauge transformation $\mathcal{G}$.  In \cite{PhysRevA.86.042318}, it is shown that without loss of generality we can define the word operators to have the form in Eq.~(\ref{eq:base_word_operator}).

The number of word operators are equal to the dimension of the code subsystem.  To encode $K$ logical states, we need $K$ word operators.  The basis state is
\begin{equation}
w'_l|S'\rangle \equiv |w'_l\rangle = |\mathbf{x}\rangle \otimes
Z^\mathbf{u}X^\mathbf{w}|\Phi_+\rangle^{\otimes c} \otimes
|\psi\rangle .
\end{equation}

The base state of an EAOCWS code doesn't include information qubits.  So we have to consider how to encode an information state $|\phi\rangle$ into a state $|\varphi'\rangle$ in the code space spanned by the states $|w_l'\rangle$.  If we assume that $|\phi\rangle$ is a $K$-dimensional system state
\[
|\phi\rangle = \sum_{l=0}^{K-1} \alpha_l |l\rangle ,
\]
we prepare the base state $|S'\rangle$ and define a unitary transformation $U_{w}$ \cite{PhysRevA.84.062321} that swaps the state $|\phi\rangle$ into the codeword:
\begin{equation}
U_{w}(|\phi\rangle\otimes |S'\rangle) = |0\rangle \otimes \sum^{K-1}_{l=0}\alpha_l |w'_l\rangle 
\equiv |0\rangle \otimes |\varphi'\rangle .
\label{Uw_encoding}
\end{equation}

$U_w$ maps a $K$-dimensional logical state onto the $K$ basis states of the code space.  After the swap operation, an encoding operation is performed on Alice's qubits.  A unitary encoding operator $U_E$ is defined by the graph state in standard form.  Each stabilizer generator has an $X$ operator on one qubit and $Z$ operators on the neighboring qubits of the graph state.  This concept doesn't change significantly in the case for EAOCWS codes.  After the unitary encoding operation $U_E$, the word gauge group for the encoded codespace is generated by following operators.  First, the word stabilizer $\mathcal{S}^s$ corresponding to Eq.~(\ref{Eq:word_stabilizer_initial}) is generated by
\begin{equation} \begin{array}{l}
X_1Z_2I \cdots IZ_n|I^{\otimes c} , \\
Z_1X_2Z_3I \cdots I|I^{\otimes c} , \\
\qquad\qquad \vdots \\
I\cdots I Z_{m-1}X_mZ_{m+1}I \cdots I|I^{\otimes c} .
\end{array}
\end{equation}
The operators that generate the group $\mathcal{S}^c$ corresponding to the $c$ ebits are
\begin{equation} \begin{array}{l}
I\cdots I Z_mX_{m+1}Z_{m+2}I \cdots I|Z_1 I\cdots I , \\
I\cdots I Z_{m+1}I \cdots I|X_1 I\cdots I , \\
\qquad\qquad\vdots \\
I\cdots I Z_{m+c-1}X_{m+c}Z_{m+c+1}I \cdots I|I\cdots I Z_c , \\
I\cdots I Z_{m+c}I \cdots I|I\cdots I X_c . \\
\end{array}
\end{equation}
The gauge group $\mathcal{S}^g$ corresponding to arbitrary transformations on subsystem $\mathcal{B}$ is generated by
\begin{equation} \begin{array}{l}
I\cdots I Z_{n-r}X_{n-r+1}Z_{n-r+2}I \cdots I|I^{\otimes c} , \\
I\cdots I Z_{n-r+1}I \cdots I|I^{\otimes c} , \\
\qquad\qquad\vdots \\
Z_1I\cdots I Z_{n-1}X_n|I^{\otimes c} , \\
I\cdots I Z_n|I^{\otimes c} .
\end{array}
\end{equation}
Therefore, the word gauge group $\mathcal{G}$ for the base state $|S\rangle$ is generated by
\begin{equation*}
\mathcal{G}=\langle \{\mathcal{S}^s,\mathcal{S}^c,\mathcal{S}^g\} \rangle .
\end{equation*}

The reason that a CWS code can be constructed from a classical binary code is that all the errors that occur in the channel are equivalent to effective errors comprising only $Z$ and $I$ operators.
By the word gauge group, a similar transformation is possible on an EAOCWS code:  all the errors occurring in the channel can be converted to errors comprising only $Z$ and $I$ operators.  For example, consider an EAOCWS code on $6$ physical qubits with $c=1$ ebit and $r=1$ gauge qubit.  For this code, the word gauge group is generated by
\begin{eqnarray*}
s_1&=&XZIIZ|I,\\ 
s_2&=&ZXZII|I,\\
s_3&=&IZXZI|I,\\
s_4&=&IIZXZ|Z,\\
s_5&=&ZIIZX|I,\\
h_1&=&IIIZI|X,\\
g_1&=&IIIIZ|I.
\end{eqnarray*}
If an error $E=IXIIX|I$ occurs on a codeword, the effective error induced from $E$ can be represented as a binary vector by Eq.~(\ref{Eq:CWS_map}):
\begin{equation}
\label{Eq:ex_binary}
Cl_G(E=IXIXI|I)=10000|1 .
\end{equation}
This binary vector is determined by applying operators $s_2$, $s_4$ and $g_1$ to the error $E$ to eliminate the $X$ operators and leave only $Z$s.

As stated above, it is assumed that the physical errors do not affect Bob's qubits.  However, as shown in Eq.~(\ref{Eq:ex_binary}), when we convert to an effective error,  there can be $Z$ operators acting on Bob's side resulting from the $h_i$ operators.  Since the word operators correspond to codewords from a binary code designed to correct this set of effective errors, this means that the word operators will also include operators that act on both Alice's and Bob's qubits, in general.  In \cite{PhysRevA.84.062321}, it was shown that $Z$ operators on Bob's side in the word operators can be removed by applying elements of the word stabilizer.  Therefore, it is possible to construct word operators that act only on Alice's side.

This is not quite the end of the story.  As we will see in our examples below, even a set of word operators that act only on Alice's side may not allow the encoding procedure described in (\ref{Uw_encoding}) to be carried out.  Since the operator $U_w$ is a generalized swap operation, it can at most encode a state whose dimension is no greater than that of the set of ancillas in the initial base state.  If two word operators act on the base state in a way that is indistinguishable on Alice's side alone, they cannot both be used in the encoding circuit.  Therefore, we may be able to use only a subset of the word operators.  We will see below how this can arise.

%======================================================
\subsection{Examples}

We now show some examples of EAOCWS codes based on the construction introduced in the previous section.  All these codes use a base state based on the ring graph, and the particular binary codes and word operators were found by numerical search.

\begin{table*}[!htdp]
\caption{\label{T:d=3}EAOCWS codes with $d=3$}
\center
\begin{ruledtabular}
\begin{tabular}{c|ccccc}
%\hline
$r\setminus c$ & 1 & 2 & 3 & 4 & 5\\
\hline\hline
1& - & [[5,2,1,3;2]] & [[5,1,1,3;3]] & - & - \\
2& - & - &  -  & - & -\\
3& - & - & - & - & -\\
\hline
1 & [[6,1,1,3;1]] & [[6,2,1,3;2]] & ((6,4,1,3;3)) & [[6,1,1,3;4]] & -\\
2 & - & [[6,2,2,3;2]] & [[6,1,2,3;3]] & -  & -\\
3 & - & - & - & - & -\\
4 & - & - & - & - & -\\
\hline
1 & ((7,4,1,3;1)) & ((7,8,1,3;2)) & ((7,7,1,3;3)) & [[7,2,1,3;4]] & [[7,1,1,3;5]]\\
2 & [[7,1,2,3;1]] & [[7,2,2,3;2]] & [[7,2,2,3;3]] & [[7,1,2,3;4]] & - \\
3 & - & [[7,2,3,3;2]] & [[7,1,3,3;3]] &  - & - \\
4 & - & - & - & - & -\\
5 & - & - & - & - & -\\
%\hline
\end{tabular}
\end{ruledtabular}
\end{table*}

\begin{table*}[!htdp]
\caption{\label{T:d=5}EAOCWS codes with  $d=5$}
\center
\begin{ruledtabular}
\begin{tabular}{c|cccccc}
%\hline
$r\setminus c$ & 1 & 2 & 3 & 4 & 5 & 6 \\
\hline\hline
1 & - & - & - & [[7,2,1,5;4]] & [[7,1,1,5;5]] & -\\
2 & - & - & - & - & - & -\\
3 & - & - & - & -& - & -\\
4 & - & - & - & - & -& -\\
5 & - & - & - & - & -& -\\
%\hline
\end{tabular}
\end{ruledtabular}
\end{table*}

A ring graph $G$ consists of $n$ vertices arranged in a closed loop, so each vertex has exactly two neighbors.  From this ring graph, the base state of the EAOCWS can be defined.  For example, suppose an EAOCWS code is defined on $9$ physical qubit state with $c=3$ and $r=1$.  The initial base state of this code (before encoding) is
\begin{equation}
|S_b\rangle= |0\rangle|\Phi_+\rangle^{\otimes 3}|\psi\rangle ,
\label{initial_base_state_1}
\end{equation}
where $|\psi\rangle$ is an arbitrary $1$-qubit state.  The word gauge group of this code is generated by the operators which is given by
\begin{eqnarray*}
\left.\begin{array}{ll}
s_1=XZIIIZ|III , &  \\
s_2=ZXZIII|III , & \\
s_3=IZXZII|ZII , & h_1=IIZIII|XII , \\
s_4=IIZXZI|IZI , &  h_2=IIIZII|IXI , \\
s_5=IIIZXZI|IIZ , &  h_3=IIIIZI|IIX , \\
s_6=ZIIIZX|III , & g_1=IIIIIZ|III .
\end{array}\right.
\end{eqnarray*}
Using this group, all single errors that occur in the channel are mapped onto induced errors consisting only of $Z$ and $I$ operators.  First, by applying $s_i$ to all single-qubit errors, all possible $X$, $Y$ and $Z$ errors can be represented as the following induced errors:                                                                                                                                                                                                                                                                                                                                                                                                                                                                                                                                                                                                                                                                                                                                                                                                                                                                                                                                                                                                                                                                                                                                                                                                                                                                                                                                                                                                                                                                                                                                                                                                                                                                                                                                                                                                               
\begin{eqnarray*}
\left.\begin{array}{ccc}
ZIIIII|III & IZIIIZ|III & ZZIIIZ|III \\
IZIIII|III & ZIZIII|III & ZZZIII|III\\
IIZIII|III & IZIZII|ZII & IZZZII|ZII\\
IIIZII|III & IIZIZI|IZI & IIZZZI|IZI \\
IIIIZI|III & IIIZIZ|IIZ & IIIZZZ|IIZ \\
IIIIIZ|III & ZIIIZI|III & ZIIIZZ|III .
\end{array}\right.
\end{eqnarray*}
After that, we apply $g_i$ to the induced errors to remove any $Z$ operators located on the $6$th qubit:
\begin{eqnarray*}
\left.\begin{array}{ccccc}
ZIIIII|III & IZIIII|III & ZZIIII|III \\
IZIIII|III & ZIZIII|III & ZZZIII|III\\
IIZIII|III & IZIZII|ZII & IZZZII|ZII\\
IIIZII|III & IIZIZI|IZI & IIZZZI|IZI \\
IIIIZI|III & IIIZII|IIZ & IIIZZI|IIZ \\
IIIIII|III & ZIIIZI|III & ZIIIZI|III .
\end{array}\right.
\end{eqnarray*}
So, the set of effective errors are as follows:
\begin{eqnarray*}
\left.\begin{array}{ccc}
ZIIIII|III  & ZIZIII|III & ZZIIII|III \\
IZIIII|III & ZIIIZI|III & ZZZIII|III  \\
 IIZIII|III & IZIZII|ZII & IZZZII|ZII \\
IIIZII|III  & IIZIZI|IZI & IIZZZI|IZI \\
 IIIIZI|III & IIIZII|IIZ & IIIZZI|IIZ .
\end{array}\right.
\end{eqnarray*}
A classical binary code that can correct binary errors corresponding to these effective errors has codewords
\begin{eqnarray*}
\label{Eq:error_pattern}
000000|000\phantom{1}110100|010\phantom{1}110100|101\phantom{1}110100|110\\
111100|011\phantom{1}000010|101\phantom{1}010100|111\phantom{1}101000|100 .
\end{eqnarray*}
From the above binary vectors, the word operators of this code are $w_l=Z^{\mathbf{c}_l}$:
\begin{eqnarray*}
IIIIII|III\phantom{1}ZZIZII|IZI\phantom{1}ZZIZII|ZIZ\phantom{1}ZZIZII|ZZI\\
ZZZZII|IZZ\phantom{1}IIIIZI|ZIZ\phantom{1}IZIZII|ZZZ\phantom{1}ZIZIII|ZII .
\end{eqnarray*}
The $Z$ operators on Bob's side of the word operators are eliminated by applying word stabilizer elements:
\begin{eqnarray*}
IIIIII|III\phantom{1}ZZZYZI|III\phantom{1}ZIXZXZ|III\phantom{1}ZIYXZI|III\\
ZZIXYZ|III\phantom{1}IZXIYZ|III\phantom{1}IIYYYZ|III\phantom{1}ZZYZII|III .
\end{eqnarray*}
From these operators we can find a set of initial word operators that act on Alice's side before applying the unitary $U_E$:
\begin{equation*}
\left.\begin{array}{cccc}
IIIIII  & XXIYII & XXZXZI & XXZYII\\
 XXXYZI & IIZIYI &  IXZYZI & XIYIII .
\end{array}\right.
\end{equation*}
But now we run into the encoding limitation that was mentioned above.  Consider the two operators $IIIIII$ and $IIZIYI$ acting on the initial base state $|S_b\rangle$ in Eq.~(\ref{initial_base_state_1}).  These operators do produce two orthogonal states.  But these states cannot be reliably distinguished by any measurement on Alice's side alone.  This, in turn, means that it is impossible to define the ``swap'' operator $U_w$ in (\ref{Uw_encoding}).  Our code cannot include both of these codewords.  (Note that these operators could be used to encode {\it classical} information, in a manner analogous to superdense coding.)

The reason that these operators produce locally indistinguishable states is because both of them have a same operator (in this case, `$II$') acting on the two ancilla qubits $|00\rangle$.  Similarly, the states produced by the word operators
\[
XXIYII\phantom{1}XXZXZI\phantom{1}XXZYII\phantom{1}XXXYZI
\]
cannot be distinguished on Alice's side alone, because they all act with the same operator $XX$ on the two ancilla qubits.  Therefore, only one of these four word operators can be used, and only one of the corresponding codewords can be included.

Therefore, the initial word operators of this code---that produce an orthonormal basis for the code space and can be encoded locally by Alice---are
\[
IIIIIIIII\phantom{1}IXZYZI\phantom{1}XIYIII\phantom{1}XXIYII,
\]
and the code space is 4-dimensional.  So, this code is a nonadditive ((6,4,1,3;3)) EAOCWS code.

By a procedure like that above, we were able to construct a number of both additive and nonadditive codes in the EAOCWS framework. Table \ref{T:d=3} shows the parameters of some codes with minimum distance $d=3$, and Table \ref{T:d=5} shows the parameters of two EAOCWS codes with $d=5$.

\section{Conclusions}
\label{Sec:Conclusions}

We have presented a unified method to construct both additive and nonadditive EAQECCs on subsystems.  Our construction is based on the CWS framework, which can be specified by a graph topology and a classical binary code.  Using shared ebits between the sender and the receiver, the code rate and distance is improved.

Because of the word stabilizer elements corresponding to the shared ebits, the induced errors can have  $Z$ operators on Bob's side.  We can find equivalent word operators---including more than just $Z$ operators---that act only on Alice's side. This is necessary for the encoding operation to be possible.  However, this requirement means that not all possible word operators may be included in the code; word operators that transform the base state to states that are locally indistinguishable on Alice's side cannot all be included.  In addition, the word gauge operators can remove some of the $Z$ operators in the effective errors,  so the weight of the induced errors is reduced.

Finally, we showed an example of how to construct an EAOCWS code, and gave the parameters of several codes in the EAOCWS  framework.  These codes were found by numerical search, and include CWS codes with distance $d$ greater than 3.

\begin{acknowledgments}
TAB would like to thank Ching-Yi Lai and Mark Wilde for useful conversations.  TAB acknowledges financial support from NSF Grant CCF-0830801.
JS is grateful to Il~Kwon Sohn and Byungkyu Ahn for useful conversations.
This research was supported by Basic Science Research Program through the National Research Foundation of Korea(NRF) funded by the Ministry of Education, Science and Technology(2012-0025328)
\end{acknowledgments}

\bibliography{EA_Operator_CWS}

\end{document}